\documentclass[12pt]{iopart}
\usepackage{graphicx}
\begin{document}

\title[Future Experiments Summary]{Future Experiments in Relativistic Heavy Ion Collisions}

\author{R Bellwied}

\address{Physics Department, Wayne State University, 666 West Hancock, Detroit, MI 48201, USA}
\ead{bellwied@physics.wayne.edu}
\begin{abstract}
The measurements at RHIC have revealed a new state of matter, which
needs to be further characterized in order to better understand its
implications for the early evolution of the universe and QCD. I will
show that, in the near future, complementary key measurements can be
performed at RHIC, LHC, and FAIR. I will focus on results than can
be obtained using identified particles, a probe which has been the
basis for this conference over the past three decades. The
sophisticated detectors, built and planned, for all three
accelerator facilities enable us to measure leptons, photons, muons
as well as hadrons and resonances of all flavors almost equally
well, which makes these experiments unprecedented precision tools
for the comprehensive understanding of the physics of the early
universe.
\end{abstract}


\section{Introduction: The lessons from RHIC}

Over the past few years the measurements of anisotropic particle
flow and jet quenching at RHIC have revealed a deconfined state of
matter at high temperature and partonic density, which is
characterized best as a near perfect fluid, i.e. a collective state
with an extremely low ratio of shear viscosity to entropy. Recently
the experiments at RHIC were able to experimentally verify the
original conjecture of a state near the quantum limit through
several independent measurements of the $\eta$/s ratio. Fig.1 shows
a summary of calculations based on $<$p$_{T}$$>$-fluctuation, light
and heavy quark elliptic flow, and quenching measurements
\cite{gavin, romatschke, lacey, phenix-vis}. These calculations are
still model dependent, but it is intriguing to recognize that, if
the initial conditions assumed in the models are correct, the new
state of matter is not well described by either perturbative QCD or
any hadronic model.

\begin{figure} [!h]
\centering
\includegraphics[width=3.5in, bb=0 0 400 400]{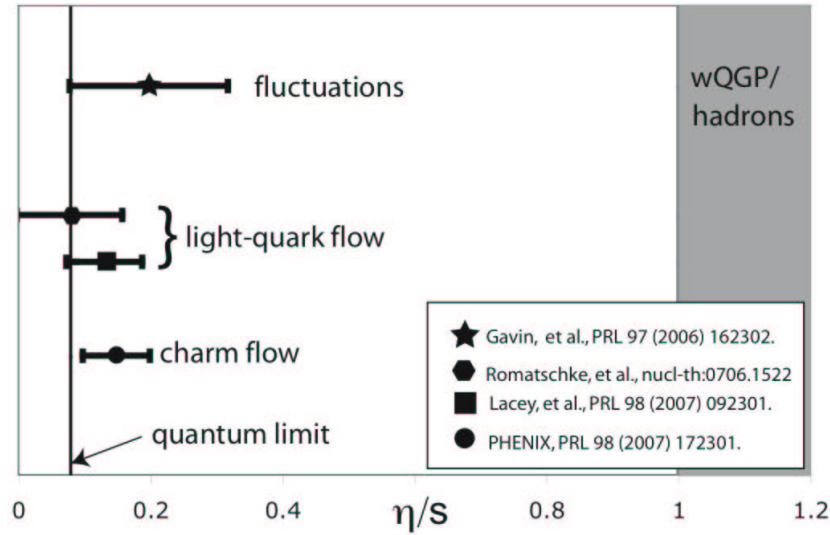}
\caption{\noindent Summary of recent model dependent determinations
of the shear viscosity over entropy ratio ($\eta$/s) based on
measurements from STAR and PHENIX.}
\end{figure}

This led to the definition of the sQGP, a strongly coupled Quark
Gluon Plasma, instead of the weak coupling phase we expected from
lattice QCD at sufficiently high initial temperature. Since then
recent lattice calculations and their comparison to hard thermal
loop calculations have revealed that the conditions at RHIC are not
sufficient to reach the weakly coupled limit, but that at LHC
energies the system likely approaches the perturbative QCD regime
\cite{lattice}. Fig.2a shows, according to Hirano and Gyulassy
\cite{hir-gyu} how the $\eta$/s ratio behaves as an order parameter
around T$_{c}$, and Fig.2b speculates on how the difference between
LHC and RHIC could be viewed in the QCD phase diagram.

\begin{figure}
\vspace{-1.cm}
\begin{minipage}{20pc}
\begin{center}
\includegraphics[width=3.5in, bb=0 -100 400 400]{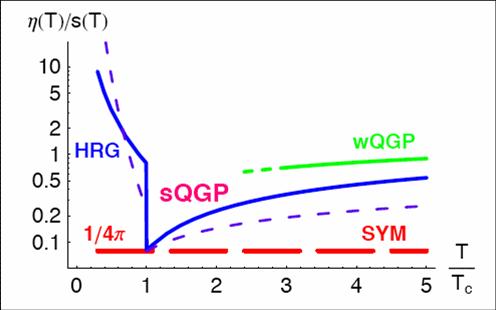}
\end{center}
\end{minipage}
\begin{minipage}{20pc}
\begin{center}
\includegraphics[width=3.8in,bb=0 0 400 400]{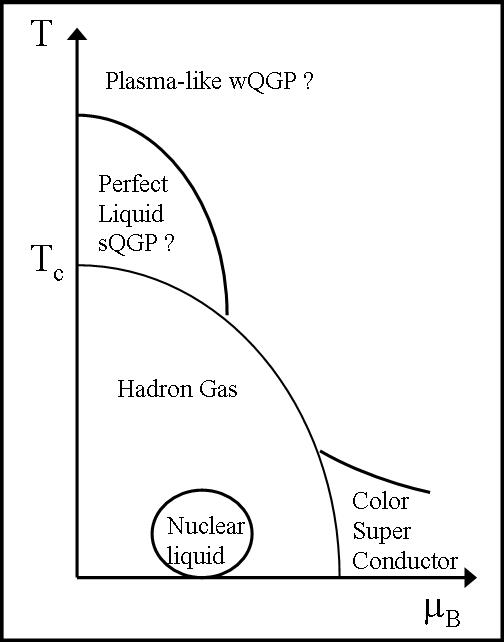}
\end{center}
\end{minipage}
\caption{a.) Behavior of $\eta$/s as a function of temperature near
T$_{c}$ according to Hirano and Gyulassy \cite{hir-gyu}, b.) New QCD
phase diagram based on latest RHIC results.}
\end{figure}

This unexpected behavior of the deconfined matter at RHIC poses
several question which drive the physics program of heavy ion
collisions to lower and higher collision energies. In my opinion
these questions are: 1.What are the degrees of freedom in the
partonic liquid ? 2. What are the thermodynamic properties (e.g.
equation of state, thermalization mechanism, response to energy
deposition) of the sQGP ? 3. What are the details of the QCD phase
transition from a strongly coupled partonic system to a hadronic gas
(hadronization mechanism) ? 4. Does the strong coupling phase change
when the initial conditions change (higher energy, lower x) ? 5.
What is the order of the phase transition and is there critical
behavior in the QCD phase diagram ? 6. Is there any experimental
evidence for chiral symmetry restoration ?

The last point was described by Ioffe \cite{yofe} in a recent
article as the remaining unresolved mystery of QCD, and he
rightfully pointed out that heavy ion collisions are the only tool
to determine the chiral transition experimentally.

Many of these questions, although asked in the narrow frame of our
field, have profound implications for the evolution of the universe.
Schwarz has summarized the physics of the first second of the
universe in a recent paper \cite{schwarz} and concluded that the QCD
phase transition might be central not only in understanding the
baryon-antibaryon asymmetry but also the generation of dark matter
in the universe. A similar conclusion was drawn by Kharzeev and
Zhvetnitsky recently \cite{kharzeev}. Certainly the theory of CP
violation in the strong sector is quite speculative, but in the time
evolution of the early universe the QCD phase transition is situated
such that most dark matter candidates have formed but not yet
decoupled from the partonic matter \cite{schwarz}, and that the
baryon-antibaryon asymmetry could be driven by the specifics of the
hadronization process at T$_{c}$.

Federico has summarized the experimental results of the past years
in his contribution \cite{antinori}, so here I will only touch on
particle identified measurements that are of particular relevance to
future experiments.

The particle identified elliptic flow and quenching measurements
have revealed a unique scaling in the intermediate transverse
momentum region \cite{v2-summary, salur}. This so-called constituent
quark scaling can be interpreted as evidence for not only
deconfinement, but also quark recombination. It is interesting to
note that the scaling parameter could be defined as the number of
constituent quarks or the number of valence quarks. This shows that
the relevant degree of freedom is not well defined. Popular
recombination models \cite{bass, greco} have taken the approach of
using thermalized constituent quarks, with a well defined mass, to
describe not only the flow data but also the unexpected high baryon
to meson (B/M) ratio at intermediate p$_{T}$ in heavy ion
collisions. A hump in the proper p$_{T}$-range in the B/M ratio
already exists in pp data \cite{star-str} and can be explained by a
di-quark suppression factor for baryon formation, but this mechanism
does not allow the baryon yield to exceed the meson yield, whereas
quark coalescence explicitly enhances the baryon yield over the
meson yield in the intermediate p$_{T}$-range. This difference
between pp and AA data very much suggests a melting of any
multi-quark structure which might have existed in pp in favor of a
system of deconfined, possibly thermal, quarks in AA collisions.

Extensions to the simple inclusive B/M ratio measurements in heavy
ion collisions have been recently performed by the STAR
collaboration and were shown at this meeting. Here B/M ratios were
determined in structures which appeared in high momentum
two-particle correlation measurements. Fig.3a shows a comparison of
B/M ratios in same-side and away-side jet cones triggered by a high
momentum charged particle \cite{zuo}, Fig.3b shows a comparison of
B/M ratios in the same side jet cone and the same-side long-range
correlation ridge as measured by STAR \cite{bielcikova}. In both
cases it seems that inside the unquenched jet the B/M ratio is
consistent with expectations from fragmentation models, whereas the
in-medium response to the traversing jets, either in the same side
ridge or the away side cone, leads to a ratio that is better
described by the recombination scenario.

\begin{figure}
\vspace{-1.cm} \hspace{0.5cm}
\includegraphics[width=3.5in, bb=0 0 400 400]{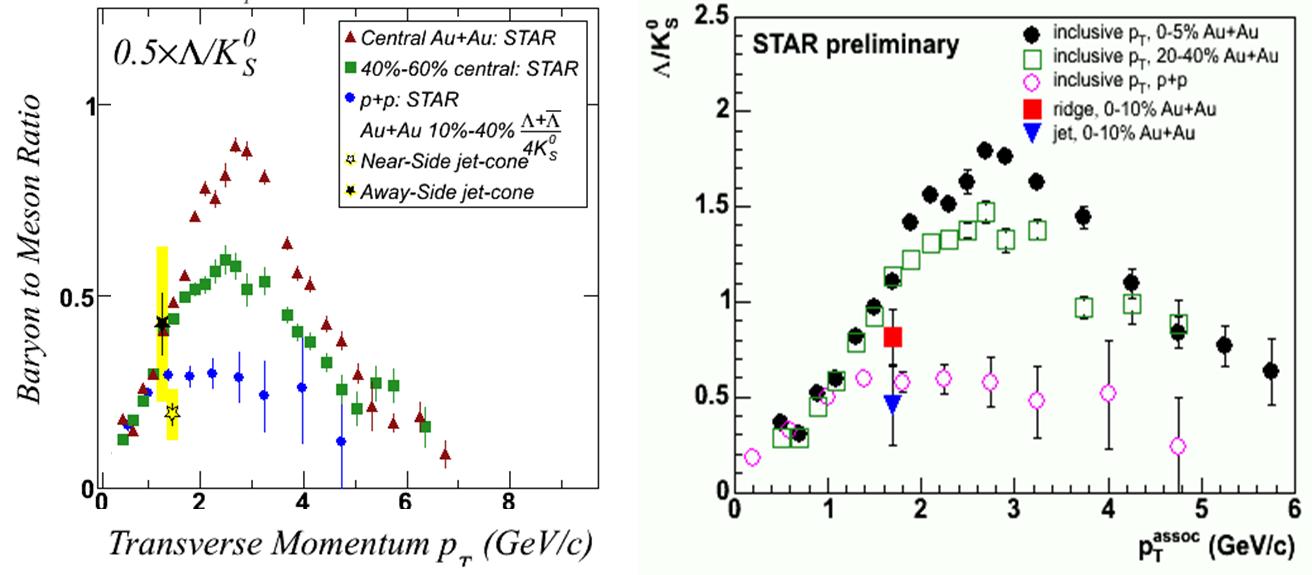} \label{fig:4}
\caption{ STAR measurements of the inclusive $\Lambda$/$K^{0}_{s}$
ratio as a function of centrality and transverse momentum, a.)
compared to ratios measured in the same-side and away-side jet
region from triggered two-particle correlations \cite{zuo}, b.)
compared to ratios measured in the same-side jet and the long-range
correlation ridge region from triggered two-particle correlations
\cite{bielcikova}.}
\end{figure}

The recent measurement of the nuclear suppression factor and the
elliptic flow for D-mesons, based on electrons from the
semi-leptonic decay of the heavy mesons
\cite{heavy-data1,heavy-data2,heavy-data3}, allows us to determine
the applicability of partonic recombination for heavy quarks, and
early results seem to indicate that both, the R$_{AA}$ and the v2
measurements, can only be explained if one assumes identical
p$_{T}$-dependencies for the flow and the quenching of light and
heavy quarks as is shown in Fig.4a for R$_{AA}$ \cite{mischke} and
in Fig.4b for v2 \cite{v2-summary}. In other words, there is a
remarkable lack of quark mass dependence in the scaling parameter.
At some high momentum the mass of the bare or constituent heavy
quark might be negligible, but this should not be the case for the
intermediate momenta measured here. Many models, as shown in Fig.4a,
have been proposed to address these measurements. The most
successful of these models try to give the heavy quark a special
status, by postulating either the survival of heavy quark resonant
states above T$_{c}$ \cite{rapp,rapp2} or the reduced formation time
of heavy quark hadrons from the partonic phase \cite{vitev}.

\begin{figure}
\begin{minipage}{20pc}
\begin{center}
\includegraphics[width=3.5in, bb=0 0 400 400]{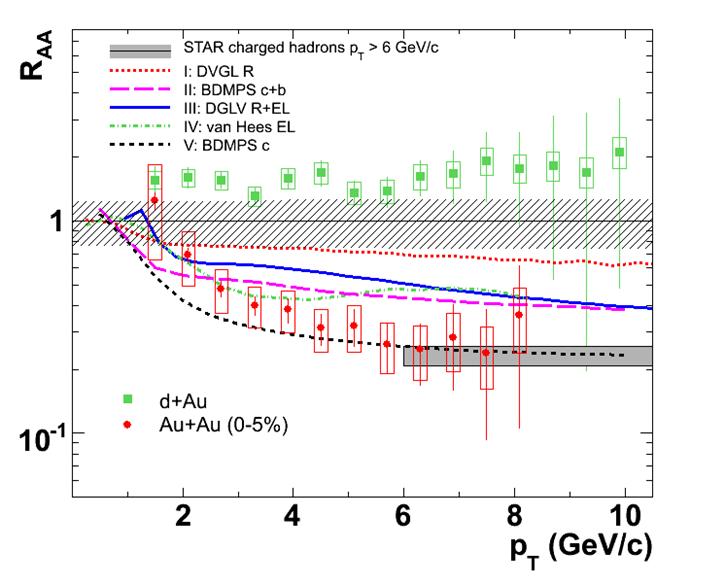} \label{fig:6}
\end{center}
\end{minipage}\hspace{1pc}%
\begin{minipage}{20pc}
\begin{center}
\includegraphics[width=3.5in, bb= 0 0 400 400]{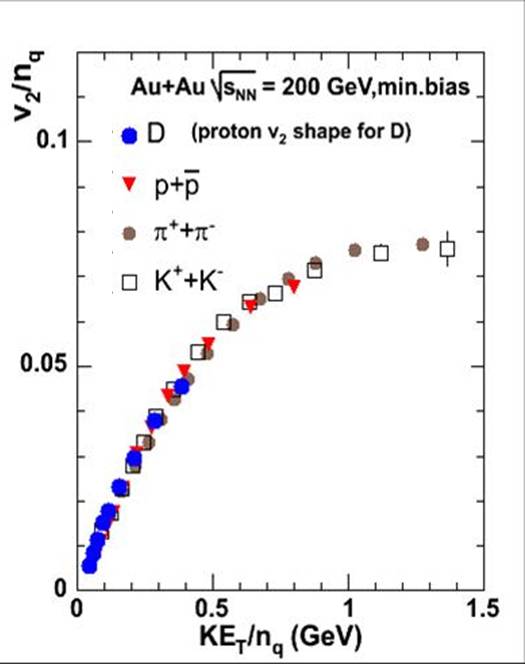} \label{fig:7}
\end{center}
\end{minipage}
\caption{a.) Nuclear suppression factor for non-photonic single
electron spectra in semi-central Au+Au collisions at RHIC compared
to the R$_{AA}$ for charged hadrons (i.e. light quark suppression)
and various models \cite{mischke}. b.) Elliptic flow (non-scaled and
n$_{q}$ scaled) for non-photonic single electron spectra in
semi-central Au+Au collisions at RHIC compared to light quark
hadrons \cite{v2-summary}.}
\end{figure}

The near identical p$_{T}$-dependence of the v2 and the quark energy
loss of light and heavy quarks is very striking, though, and might
require a much more fundamental explanation. A detailed measurement
of reconstructed D-mesons and B-mesons is sorely needed to remove
the ambiguities in the semi-leptonic measurements, and future
measurements of high momentum heavy flavor mesons and baryons should
answer the question whether the liquid phase above the critical
temperature requires indeed a special degree of freedom to describe
all features of hadronization from a dense deconfined medium. Both
STAR and PHENIX have proposed extensive upgrades to their vertexing
and tracking capabilities in order to be able to directly
reconstruct the displaced vertices of the hadronic decay channels of
heavy quark hadrons \cite{hft, phenix}.

\section{The FAIR/RHIC low energy program: Is it critical ?}

Besides the very detailed and strong evidence for deconfinement at
RHIC energies, the data reveal a surprising lack of evidence for
chiral symmetry restoration. Vector meson and resonance measurements
have been performed to new levels of precision at RHIC, in
particular in the sector of heavy hadronic resonances, but the
measurements are mostly used to determine the lifetime of the
produced partonic and hadronic systems through detailed mapping of
re-interaction probabilities \cite{markert}, because the properties
of the resonances, such as mass, width, and branching ratios are
generally in very good agreement with the particle data group
references. There are small variations in mass and width for certain
resonances as a function of their momentum, but they have been
measured consistently in pp, dA, and AA collisions
\cite{fachini1,fachini2} and therefore should not be attributed to
chiral symmetry restoration. Recently PHENIX has shown results that
might indicate behavior similar to the NA45 and NA60 low mass
di-lepton measurements \cite{phenix-dil}, but whether this is
evidence for medium modification of vector mesons remains to be
seen. It seems that either our measurements are not sensitive to
chiral symmetry restoration or that the chiral transition might
indeed decouple from deconfinement, which is in disagreement with
lattice QCD calculations. A very high luminosity program at FAIR
should enable more detailed measurements of medium modification, in
particular for chiral partners in the heavy quark sector. Not only
will the lifetime of the dense hadronic phase be considerably longer
than at RHIC, but, although open charm production is at threshold at
FAIR energies, the yield of open charm obtained in a 25 week run at
CBM is about an order magnitude larger than the yield STAR obtains
over the same period of time \cite{senger}. Detailed measurements,
not only of chiral symmetry restoration, but also particle
identified elliptic flow, radial flow and jet quenching might
therefore be possible albeit at a slightly lower p$_{T}$-range. The
key purpose of these measurements should be to map out the
disappearance of the main sQGP signatures, such as quark scaling,
hydro scaling, and high p$_{T}$ suppression in order to determine
the exact phase transition point in the QCD phase diagram. These
results will complement the thrust of the CBM program and also the
low-energy RHIC program, which focusses on the search for a critical
point. In that context, STAR is upgrading its detector to include a
full azimuth TOF device to allow the measurements of particle
identified yield ratios on an event by event basis. Besides
addressing the remarkable structure in the integrated
K$^{+}$/$\pi^{+}$ ratio as function of collision energy, as measured
by NA49 \cite{blume}, the event-by-event K/$\pi$ ratio will together
with $<$p$_{T}$$>$, charge, and baryon number measurements be
sensitive to critical dynamic fluctuations.

\section{The LHC program: Vaporizing the Quark Soup}

If the strong coupling strength reaches the weak limit at around 3
T$_{c}$, then it is very likely that at LHC energies we will indeed
reach the plasma, rather than liquid, phase which was originally
anticipated for RHIC energies. This phase will only exist for a few
fm/c and then has to de-excite through the strong coupling phase to
the hadronization surface, but the question arises whether the weak
coupling during the early times might lead to any measurable
features. It is likely that the hadronization mechanism is not
affected, but collective phenomena which are supposed to develop
early, such as elliptic flow, might be reduced by a weak coupling
phase. One can also speculate that the system might be more dilute
when it enters the strong coupling regime, and therefore exhibits
less of a collectivity.

On the other hand, the partonic system is expected to live longer,
and estimates by Eskola et al. \cite{eskola} show that the
applicability of hydrodynamics might extend to higher p$_{T}$ which
means that the thermal bulk will begin to populate the intermediate
p$_{T}$ range. At the same time it is likely that the quark
coalescence mechanism will push out to higher p$_{T}$ simply because
the thermal partons will carry more energy at LHC energies
\cite{bass2}. Finally the increase in jet cross section will lead to
enhanced particle production in the intermediate p$_{T}$ range
through jet quenching \cite{borghini}. In that context, hybrid
models which allow the recombination of thermal partons with hard
fragmentation partons claim to predict the particle spectrum at the
LHC over a wide momentum range (2-20 GeV/c) \cite{hwa2}. Clearly,
although the intermediate momentum range is populated much stronger
at the LHC than at RHIC, it will be challenging to disentangle all
these different contributions to hadron production. On the other
hand the large statistics will allow us to study the transition from
bulk matter production to non-thermal contributions at higher
momentum in a systematic way at the LHC. In particular ALICE is well
suited for hadronization studies through its superior particle
identification capabilities.

Certain predictions made at this conference have revealed a great
sensitivity to the properties of the plasma at the LHC. I will only
list a few that I think are of special relevance to the theme of
this conference.

The question of equilibrium vs. non-equilibrium at the time of
hadronization has still not been settled through the RHIC
measurements, but at the LHC the possible oversaturation of
strangeness in a non-equilibrium approach leads to a novel effect as
shown by Rafelski \cite{raf}, namely the binding of charm by strange
quarks into rare heavy states, such as the D$_{s}$ and the $\Xi_{c}$

In the heavy sector the ratio of charm quenching over bottom
quenching is expected to approach unity at sufficiently high
transverse momentum (p$_{T}$ $>$ 50 GeV/c) if pQCD is correct,
whereas a AdS/CFT based calculation shows no sensitivity to the
change in collision energy and predicts a constant quenching ratio
for c/b independent of the particle momenta as shown by Horowitz
\cite{horowitz}. A similar cross-check is given by comparing muons
from W decays to muons from B-meson decays. The muons from W decays
are effectively medium blind, whereas the B-meson muons should
exhibit the effect of jet quenching. Dainese \cite{dainese} showed
that the crossing point for W/B muons in transverse momentum will be
sensitive to the level of partonic energy loss for B-mesons in
medium.

The measurement of all onium states, and in particular separate
measurements of all Y states will enable a comprehensive look at
color screening in the medium. The J/$\psi$ measurement holds
additional information, because the abundant production of c-cbar
pairs at the LHC increases the probability of partonic recombination
of J/$\psi$ to offset the screening effect. Therefore recombination
models have the J/$\psi$ yield actually rise from RHIC to the LHC
\cite{pbm}. It was interesting to note at this conference, though,
that hadronic recombination models, which favor the coalescence of
charm bound in D-mesons during the co-moving hadronic phase, yield
the same J/$\psi$ enhancement than the partonic models
\cite{bravina, lynnuk}.

Finally, the measurement of resonances in jets as proposed by
Markert \cite{markert2} might hold great promise at a more
differential probe of chiral symmetry restoration in the medium. The
idea here is based on the possibility that heavy high momentum
resonances are formed earlier in the mixed phase than the bulk
pions. In this case the resonances might undergo medium
modifications in the partonic medium but carry too high a momentum
to re-interact hadronically after bulk hadronization. It is
intriguing to recognize that the reference measurement of
hadronically interacting resonances could be made in the same event,
simply by comparing resonances correlated to the jet axis with off
axis, or uncorrelated, resonances. STAR will attempt such a
measurement but only the increased yield and momentum reach, as well
as the longer lifetime of the partonic phase at the LHC is likely to
enable us to achieve the necessary conditions.

\section{Future detector upgrades: Are they necessary ?}

Possible requirements for new detector components are based on
specific shortcomings of the existing detectors for certain key
measurements. A comparison of heavy ion detectors to elementary
particle detectors reveal some of these features. For example, it
might be desirable in a heavy ion experiment to have near hermetic
coverage, i.e. detection capabilities should extend to the highest
accessible pseudo-rapidity, in particular for decay $\gamma$ and the
away-side component of high momentum triggered $\gamma$-jet events.
It would also be interesting to have high precision vertexing and
tracking out to very high momentum without having to build even
larger gas based TPC's. This can be achieved through a solid-state
tracker in a large magnetic field. One could further extend the
granularity of electromagnetic calorimetry for photon (direct,
decay, fragmentation and thermal photons) measurements, which drives
the necessity for a crystal type calorimeter and hadronic
calorimetry to distinguish the different neutral particle energy
contributions. A key benchmark for any final calorimeter design
should be that all $\Upsilon$ states and the $\chi_{c}$ can be
reconstructed in the heavy-ion environment and that the away-side
spectrum of $\gamma$-jets can be reliably measured out to a high
$\gamma$-p$_{T}$. Finally a new device should include a very high
momentum particle identification component to extend the critical
flavor dependent measurements out to high momentum. Reliable
identified singles spectra should be measured to about 25 GeV/c in
order to be able to distinguish all the different contributions to
the spectrum.

These are simple hardware extensions based on the existing
limitations of the big three heavy ion detectors (STAR,PHENIX,
ALICE), but they also can be easily linked to interesting new
measurements required to fully characterize the partonic phase at
RHIC and the LHC. It is for the next generation of heavy ion
physicists to decide whether all these improvements can be realized
within a reasonable budget or whether targeted upgrades to existing
devices will be sufficient.

\section{Conclusions: The Future is bright}

The future of relativistic heavy ion physics is very bright. The
discovery of the sQGP at RHIC has shown that a new phase of matter
has existed above the critical temperature during the early
evolution of the universe. It is encumbant upon us to characterize
this phase in great detail and to explore the likely far-reaching
implications for the evolution of matter during the first second of
the universe. Both the low energy program at RHIC/FAIR as well as
the final frontier program at the LHC are absolutely crucial for
understanding this new phase of matter. In addition it might turn
out that the RHIC energy constitutes a sweet spot for studying the
process of hadronization from strongly coupled partonic matter to
hadrons. Therefore all three programs will have a very strong future
ahead of them.

\vfill\eject

\end{document}